\documentclass[%
 aip,
 rsi,%
 amsmath,amssymb,
 reprint,%
]{revtex4-1}

\usepackage{graphicx}% Include figure files
\usepackage{dcolumn}% Align table columns on decimal point
\usepackage{bm}% bold math
\begin{document}

%\preprint{}

\title{A magnetically-driven piston pump for ultra-clean applications} %Title of paper



%\author{}
%\thanks{}
%\affiliation{}

\author{F.~LePort}
\altaffiliation{Now at Tesla Motors, Palo Alto CA, USA}
\author{R.~Neilson}
%\cortext[cor]{Corresponding author. Tel: +1-650-725-6342; fax: +1-650-725-6544.}
\email{rneilson@stanford.edu}
\author{P.S.~Barbeau}
\author{K.~Barry}
\author{L.~Bartoszek} 
\author{I.~Counts}
\author{J.~Davis}
\author{R.~deVoe}
\author{M.J.~Dolinski}
\author{G.~Gratta}
\author{M.~Green}
\altaffiliation{Now at University of North Carolina, Chapel Hill NC, USA}
\author{M.~Montero~D\'iez}
\author{A.R.~M\"{u}ller}
\author{K.~O'Sullivan}
\author{A.~Rivas}
\author{K.~Twelker}
\affiliation{Physics Department, Stanford University, Stanford CA, USA}
\author{B.~Aharmim}
\affiliation{Physics Department, Laurentian University, Sudbury ON, Canada}
\author{M.~Auger}
\affiliation{LHEP, Physikalisches Institut, University of Bern, Bern, Switzerland}
\author{V.~Belov}
\affiliation{Institute for Theoretical and Experimental Physics, Moscow, Russia}
\author{C.~Benitez-Medina}
\affiliation{Physics Department, Colorado State University, Fort Collins CO, USA}
\author{M.~Breidenbach}
\affiliation{SLAC National Accelerator Laboratory, Stanford CA, USA}
\author{A.~Burenkov}
\affiliation{Institute for Theoretical and Experimental Physics, Moscow, Russia}
\author{B.~Cleveland}
\affiliation{Physics Department, Laurentian University, Sudbury ON, Canada}
\author{R.~Conley}
\affiliation{SLAC National Accelerator Laboratory, Stanford CA, USA}
\author{J.~Cook}
\affiliation{Physics Department, University of Massachusetts, Amherst MA, USA}
\author{S.~Cook}
\affiliation{Physics Department, Colorado State University, Fort Collins CO, USA}
\author{W.~Craddock}
\affiliation{SLAC National Accelerator Laboratory, Stanford CA, USA}
\author{T.~Daniels}
\affiliation{Physics Department, University of Massachusetts, Amherst MA, USA}
\author{M.~Dixit}
\affiliation{Physics Department, Carleton University, Ottawa ON, Canada}
\author{A.~Dobi}
\affiliation{Physics Department, University of Maryland, College Park MD, USA}
\author{K.~Donato}
\affiliation{Physics Department, Laurentian University, Sudbury ON, Canada}
\author{W.~Fairbank~Jr.}
\affiliation{Physics Department, Colorado State University, Fort Collins CO, USA}
\author{J.~Farine}
\affiliation{Physics Department, Laurentian University, Sudbury ON, Canada}
\author{P.~Fierlinger}
\affiliation{Technical University Munich, Munich, Germany}
\author{D.~Franco}
\author{G.~Giroux}
\author{R.~Gornea}
\affiliation{LHEP, Physikalisches Institut, University of Bern, Bern, Switzerland}
\author{K.~Graham}
\author{C.~Green}
\author{C.~H\"{a}gemann}
\affiliation{Physics Department, Carleton University, Ottawa ON, Canada}
\author{C.~Hall}
\affiliation{Physics Department, University of Maryland, College Park MD, USA}
\author{K.~Hall}
\affiliation{Physics Department, Colorado State University, Fort Collins CO, USA}
\author{D.~Hallman}
\affiliation{Physics Department, Laurentian University, Sudbury ON, Canada}
\author{C.~Hargrove}
\affiliation{Physics Department, Carleton University, Ottawa ON, Canada}
\author{S.~Herrin}
\affiliation{SLAC National Accelerator Laboratory, Stanford CA, USA}
\author{M.~Hughes}
\affiliation{Department of Physics and Astronomy, University of Alabama, Tuscaloosa AL, USA}
\author{J.~Hodgson}
\affiliation{SLAC National Accelerator Laboratory, Stanford CA, USA}
\author{F.~Juget}
\affiliation{LHEP, Physikalisches Institut, University of Bern, Bern, Switzerland}
\author{L.J.~Kaufman}
\affiliation{Physics Department, Indiana University, Bloomington IN, USA}
\author{A.~Karelin}
\affiliation{Institute for Theoretical and Experimental Physics, Moscow, Russia}
\author{J.~Ku}
\affiliation{SLAC National Accelerator Laboratory, Stanford CA, USA}
\author{A.~Kuchenkov}
\affiliation{Institute for Theoretical and Experimental Physics, Moscow, Russia}
\author{K.~Kumar}
\affiliation{Physics Department, University of Massachusetts, Amherst MA, USA}
\author{D.S.~Leonard}
\affiliation{Department of Phyiscs, University of Seoul, Seoul, Korea}
\author{G.~Lutter}
\affiliation{LHEP, Physikalisches Institut, University of Bern, Bern, Switzerland}
\author{D.~Mackay}
\affiliation{SLAC National Accelerator Laboratory, Stanford CA, USA}
\author{R.~MacLellan}
\affiliation{Department of Physics and Astronomy, University of Alabama, Tuscaloosa AL, USA}
\author{M.~Marino}
\affiliation{Technical University Munich, Munich, Germany}
\author{B.~Mong}
\affiliation{Physics Department, Colorado State University, Fort Collins CO, USA}
\author{P.~Morgan}
\affiliation{Physics Department, University of Massachusetts, Amherst MA, USA}
\author{A.~Odian}
\affiliation{SLAC National Accelerator Laboratory, Stanford CA, USA}
\author{A.~Piepke}
\affiliation{Department of Physics and Astronomy, University of Alabama, Tuscaloosa AL, USA}
\author{A.~Pocar}
\affiliation{Physics Department, University of Massachusetts, Amherst MA, USA}
\author{C.Y.~Prescott}
\affiliation{SLAC National Accelerator Laboratory, Stanford CA, USA}
\author{K.~Pushkin}
\affiliation{Department of Physics and Astronomy, University of Alabama, Tuscaloosa AL, USA}
\author{E.~Rollin}
\affiliation{Physics Department, Carleton University, Ottawa ON, Canada}
\author{P.C.~Rowson}
\affiliation{SLAC National Accelerator Laboratory, Stanford CA, USA}
\author{B.~Schmoll}
\affiliation{Physics Department, University of Massachusetts, Amherst MA, USA}
\author{D.~Sinclair}
\affiliation{Physics Department, Carleton University, Ottawa ON, Canada}
\author{K.~Skarpaas}
\affiliation{SLAC National Accelerator Laboratory, Stanford CA, USA}
\author{S.~Slutsky}
\affiliation{Physics Department, University of Maryland, College Park MD, USA}
\author{V.~Stekhanov}
\affiliation{Institute for Theoretical and Experimental Physics, Moscow, Russia}
\author{V.~Strickland}
\affiliation{Physics Department, Carleton University, Ottawa ON, Canada}
\author{M.~Swift}
\affiliation{SLAC National Accelerator Laboratory, Stanford CA, USA}
\author{J.-L.~Vuilleumier}
\author{J.-M.~Vuilleumier}
\affiliation{LHEP, Physikalisches Institut, University of Bern, Bern, Switzerland}
\author{U.~Wichoski}
\affiliation{Physics Department, Laurentian University, Sudbury ON, Canada}
\author{J.~Wodin}
\author{L.~Yang}
\affiliation{SLAC National Accelerator Laboratory, Stanford CA, USA}
\author{Y.-R.~Yen}
\affiliation{Physics Department, University of Maryland, College Park MD, USA}

\collaboration{EXO Collaboration}
\noaffiliation

\date{\today}

\begin{abstract}
A magnetically driven piston pump for xenon gas recirculation is presented. The pump is designed to satisfy extreme purity and containment requirements, as is appropriate for the recirculation of isotopically enriched xenon through the purification system and large liquid xenon TPC of EXO-200. The pump, using sprung polymer gaskets, is capable of pumping more than 16 standard liters per minute (SLPM) of xenon gas with 750~torr differential pressure.

\end{abstract}

%\pacs{}% insert suggested PACS numbers in braces on next line

%Use showkeys class option if keyword

\maketitle %\maketitle must follow title, authors, abstract and \pacs

\section{Introduction}

There is often a need for forced gas flow in high-purity gas systems, requiring a reliable, ultra-clean, pump with superior containment properties. Available pump technologies, such as diaphragm and bellows pumps, can be made with clean materials. For example, the MUNU experiment \cite{amsler97} used a bellows pump and the XENON10 experiment \cite{aprile10} used a diaphragm pump. However, for these designs, a failure of a bellows or a diaphragm results in a breach of the gas system, necessitating double containment and leak detection to avoid gas loss and contamination. Even with double containment, a failure results in down time and risk of contamination of the system during the repair. A pump that is both extremely clean and robust with respect to gas containment is desirable, and is the subject of this paper.

The pump described here has been developed for xenon gas recirculation and forced condensation in the Enriched Xenon Observatory (EXO). EXO is a program aimed at building a ton-scale neutrinoless double beta ($0\nu\beta\beta$) decay \cite{avignone08} detector using xenon enriched to 80\% in the isotope $^{136}$Xe as the source and detection medium \cite{danilov00, *breidenbach01}. In particular, a version of this pump is currently operating in EXO-200---an intermediate scale (200~kg of enriched xenon, 80\% $^{136}$Xe) liquid xenon detector. The xenon gas in EXO-200 must be kept extremely clean of
electro-negative impurities and radioactive backgrounds. For optimal ($>$1~ms) electron lifetimes, the xenon must have $<$0.3~parts-per-billion O$_2$-equivalent contamination \cite{bakale76}. To meet these cleanliness requirements, EXO-200 is designed to allow for continuous recirculation of xenon (for several years of data taking) through hot zirconium purifiers\cite{saes} and a radon trap in gas phase---hence the need for a xenon pump.

The xenon pump is used in three distinct modes of operation. Firstly, prior to initial liquefaction xenon gas is recirculated through the purifiers to purge impurities from the xenon vessel and gas system. Secondly, during initial liquefaction the xenon pump forces gas though a xenon condenser for accelerated condensation. Thirdly, and most importantly, when the vessel is full of liquid, gaseous xenon evaporated by a heater is pumped through the purifiers and re-condenses in a condenser, establishing continuous purification of the xenon during detector operations.

Double containment, all-metal bellows pumps were tested but were found to be plagued by leaks after periods of operation greater than a month, even when selecting long-lifetime bellows and carefully designing the stroke for durability. The pump described here makes use of magnetic coupling between a pair of permanent magnets, one inside the process fluid (gas Xe) and the other in air outside the gas system, eliminating the need for flexible components isolating the process fluid from the atmosphere. This eliminates the possibility of contamination of the xenon gas due to component failure, limiting the effects of long term wear to a possible decrease of pumping efficiency.

The drive mechanism, located completely outside the xenon system, moves an external neodymium ring magnet\cite{extmag} back and forth over a stainless steel cylinder. Inside the cylinder a piston, containing a sealed cylindrical neodymium magnet\cite{intmag}, moves in phase with the external magnet and pumps the xenon gas. The magnets are axially magnetized with opposing polarity.

It is critical that the xenon pump itself is not a source of contamination, so it has been constructed such that the only materials exposed to xenon gas are stainless steel, titanium, and a carefully selected polymer used for the piston seals. Gas containment is insured exclusively by static, all-metal ConFlat\textsuperscript\textregistered\cite{conflat} and VCR\textsuperscript\textregistered\cite{vcr} seals. No lubricants are used inside the pump.

\section{Design and construction}

\begin{figure*}
\includegraphics{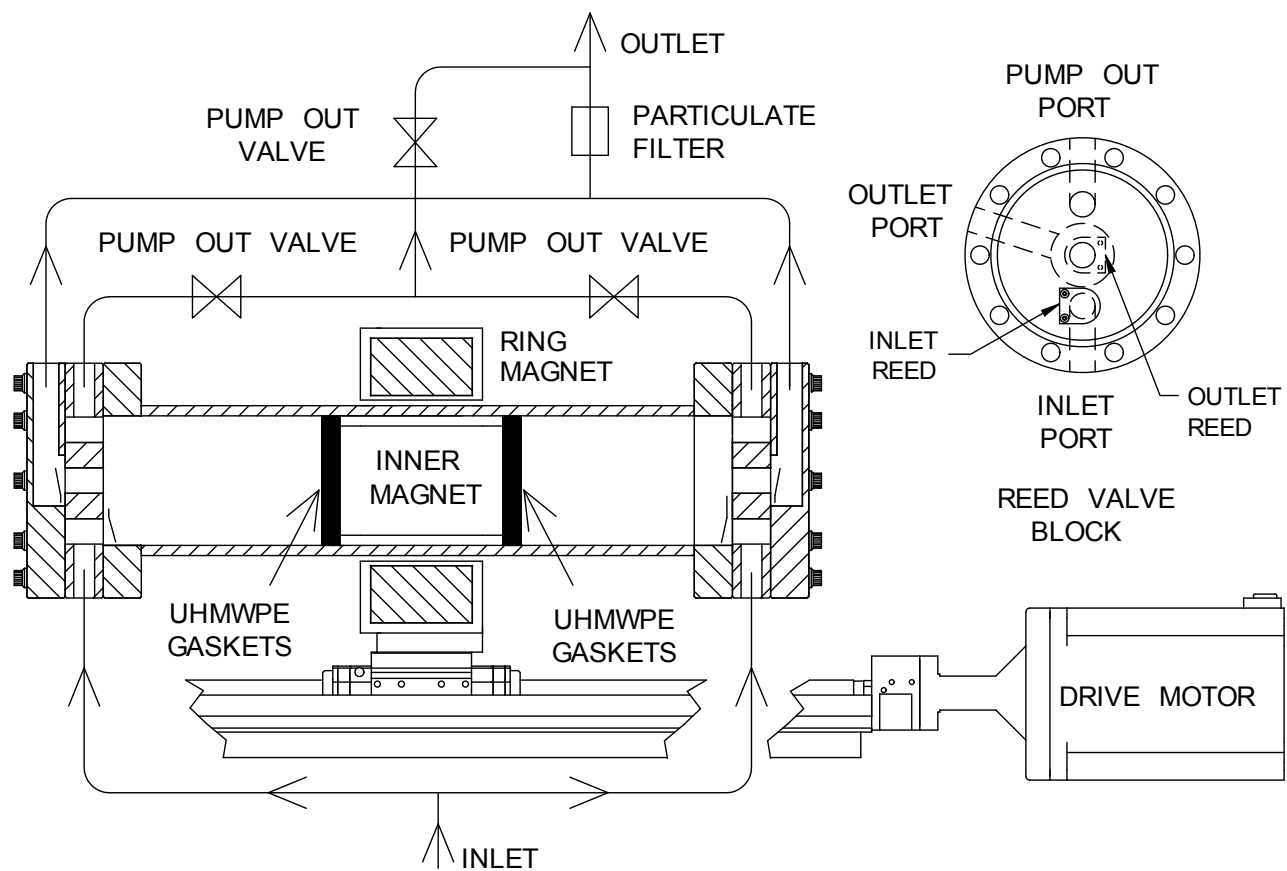}
\caption{\label{fig:schematic}Schematic diagram of the magnetically coupled piston pump. One of the reed valve blocks, made from 4-5/8\texttt{"} ConFlat\textsuperscript\textregistered\ flanges, is shown separately at the top right.}
\end{figure*}

\begin{figure}
\includegraphics{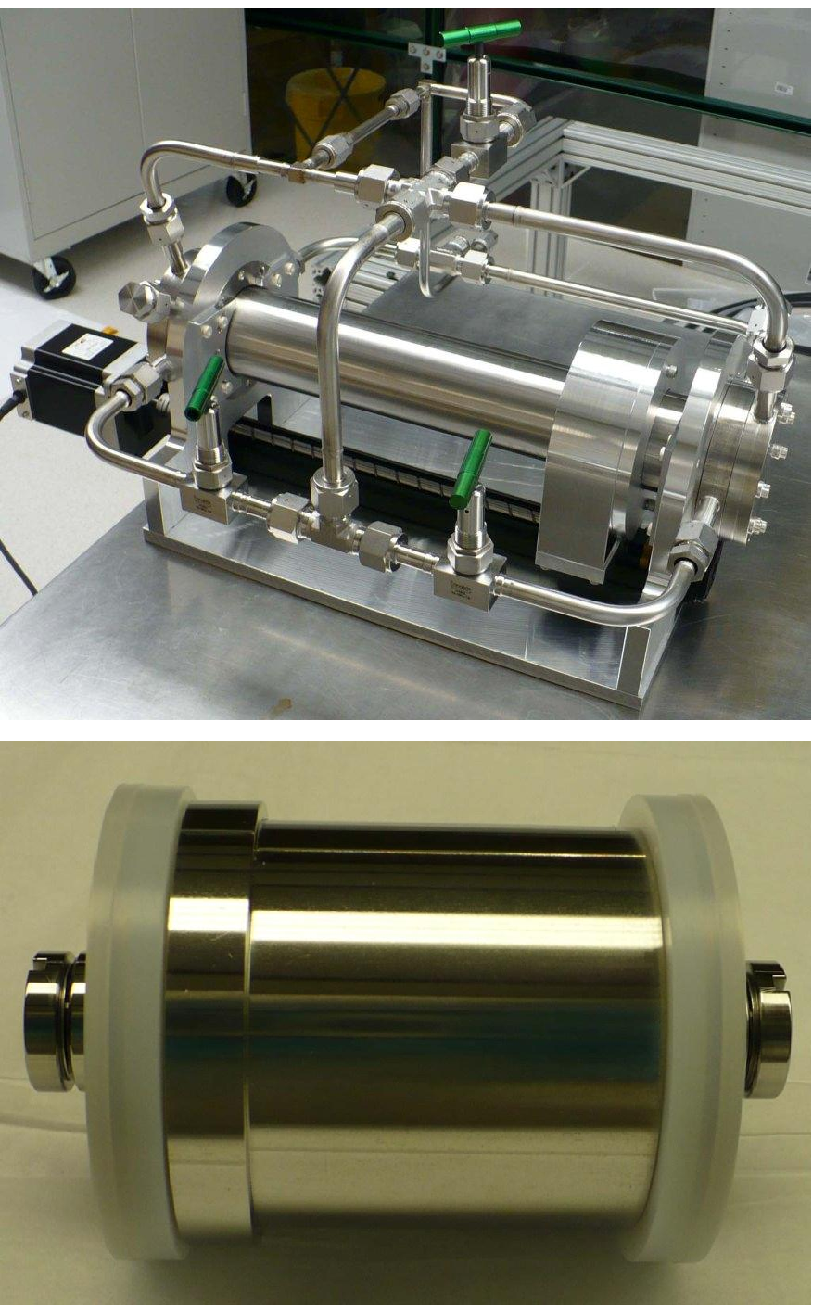}
\caption{\label{fig:photo}Top: Photograph of the assembled xenon pump. Bottom: Photograph of the piston removed from the pump.}
\end{figure}

A schematic view of the pump is shown in Figure~\ref{fig:schematic} and photographs of the pump and piston are shown in Figure~\ref{fig:photo}. The pump body is made from a honed thick-walled non-magnetic 316 stainless-steel cylinder, 40~cm long with inside diameter of 6.5~cm. The inner and outer diameters of the cylinder are chosen to tightly match the dimensions of commercial permanent magnets. The cylinder is welded onto 4-5/8\verb+"+ ConFlat\textsuperscript\textregistered\ flanges. The external ring magnet is mounted on a linear actuator\cite{nsk} driven by a stepper motor\cite{amp}. This configuration allows programmable movement of the external magnet---and thus the internal piston---up to speeds greater than 15~cm/s, and, if desired, programmable acceleration and deceleration. Optical position sensors are installed so that the drive can seek the starting location at startup and during operation to correct for drifts. 

The piston consists of a welded stainless steel cannister with the cylindrical neodymium magnet sealed inside. On both ends plastic piston gaskets are mounted with stainless steel retaining and tensioning rings. The gaskets create a seal between the piston and the honed cylinder. The cannister could not be TIG welded because of the strong magnetic field and was instead laser welded\cite{appfus}. Fasteners and tools used in the assembly are made of titanium for the same reason.

Inlet and outlet ports with reed valve assemblies, made with 50~$\mu$m thick stainless steel stock, ensure that gas flow is unidirectional. The seats of the reed valves are made by optically lapping the appropriate regions of a double-sided ConFlat\textsuperscript\textregistered\ blank in which holes were drilled. As the piston moves through the cylinder, gas is pulled in through the inlet on one end and forced out through the outlet on the other end. The two inlets and outlets are plumbed in parallel so that flow is produced in both directions of travel of the piston. Ports directly accessing the cylinder volume (bypassing the reed valves) allow for more efficient pump-out of air from the interior prior to integration in the xenon system.

A sufficiently strong coupling between the magnets is essential for the pump to achieve its required performance. If the coupling is not strong enough, the gas pressure acting on the piston will decouple the piston magnet from the external magnet, disabling the pump (although if this happens the magnets immediately re-couple on the return movement of the external magnet). The design goal was for the pump to operate up to a differential pressure of 750~torr while achieving flows in excess of 10~SLPM.

\begin{figure}
\includegraphics{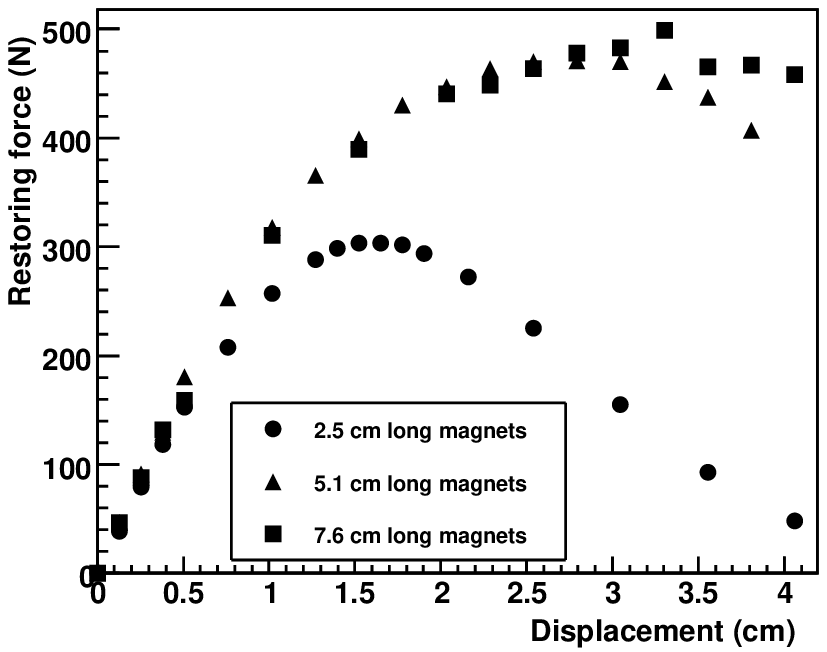}
\caption{\label{fig:maxwell}MAXWELL calculations of the restoring force acting on the piston magnet versus displacement. The maximum restoring force is the force at which the magnets will decouple.}
\end{figure}

The maximum differential pressure possible with our pump was calculated for three arrangements of commercially available magnets using the MAXWELL\cite{maxwell} software suite. For the piston magnet to remain coupled to the ring magnet, the pressure acting on the piston must be less than the maximum magnetostatic restoring force on the piston. Figure~\ref{fig:maxwell} shows the calculated restoring force for three magnet configurations. For all three configurations the inner piston magnet has a diameter of 5.1~cm, the outer ring magnet has inner diameter of 7.6~cm and outer diameter of 10.2~cm. The length of both magnets is varied from 2.5--7.6~cm between the three configurations. The calculated maximum restoring force is 310~N for the 2.5~cm long magnets, 480~N for the 5.1~cm long magnets and 490~N for the 7.6~cm long magnets. From this we calculate the maximum pressure as 730, 1130 and 1150~torr for the 2.5, 5.1 and 7.6~cm long magnets respectively.

A second design challenge is maintaining an adequate seal between the piston and pump cylinder without contaminating the xenon. While for this design a leaky seal does not compromise the integrity of the gas circuit, it does reduce the efficiency of the pump. The appropriate choice of piston ring material is critical, as is the design of the seals. The pump relies on a dynamic seal between the piston and stainless steel cylinder, and since the ultra-pure xenon gas is in contact with the piston, no lubricants can be used. To meet radiological purity requirements, only low radon out-gassing materials free of other radioactive contaminants are suitable. Friction between the pump wall and seals must be low, and the seals must be durable. Of particular concern is the possibility of particulate generated by the piston rings contaminating the xenon.

We primarily considered polymers as seal materials and have investigated polytetrafluoroethylene (PTFE), fluorinated ethylene propylene (FEP), perfluoroalkoxy (PFA), polyether ether ketone (PEEK) and ultra-high-molecular-weight polyethylene (UHMWPE).\cite{brass} Samples of these five materials were tested for wear by spinning under a calibrated pressure a part, fashioned as a tube, against an optically polished quartz plate. Wear was assessed by the amount of particulate generated in the test. For this test to be realistic both the sample material and quartz plate have to be thoroughly degreased. The spinning velocity was set to simulate the linear velocity expected in the gasket. It was found that UHMWPE exhibited the lowest wear, followed by PFA then PTFE and FEP. PEEK performed very poorly in this test. Test pistons were made out of PFA, PTFE and UHMWPE and operated in a prototype pump, confirming that the latter produces the least particulate. %These three materials all have low co-efficients of friction \cite{crc}.

\begin{figure}
\includegraphics{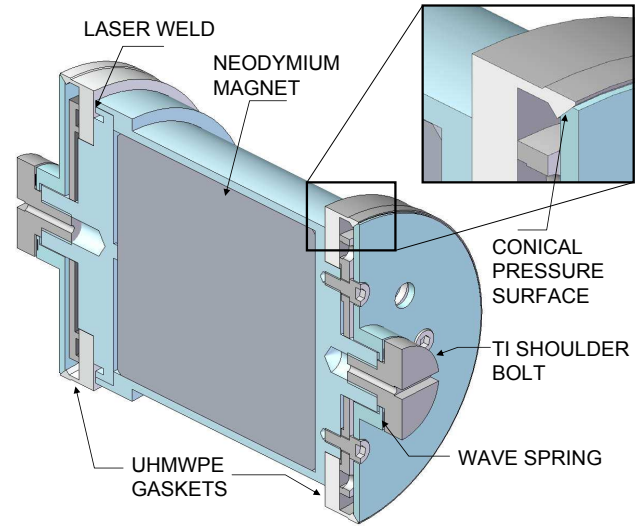}
\caption{\label{fig:piston}A view of the piston with sprung UHMWPE gaskets. Radial pressure is applied to the gaskets by axially pressing a conical stainless steel part into it by means of a wavy spring washer. The neodymium magnet is sealed inside the stainless steel cannister.}
\end{figure}

We experimented with a number of different profiles for the piston rings. The rings must seal the piston to the cylinder at speeds up to 15~cm/s and the friction must be kept to a small fraction of the magnet decoupling force. The rings must also accommodate thermal expansion from frictional heating while maintaining a constant radial pressure, and must center the piston with tight tolerance because of the unstable radial equilibrium provided by the magnetic coupling. To achieve this the plastic sealing surface is radially sprung by forcing a conical stainless steel element into it by means of an axial force applied by a spring washer. This method, illustrated in Figure~\ref{fig:piston}, also allows for adjustment of the radial force that then remains constant until the sealing surface is completely worn out. In the pump in operation at EXO-200 an axial force of 20~N is provided by the spring washer against the 45$^{\circ}$ conical surface. A simple calculation gives a radial linear pressure of 1.0~N/cm along the sealing surface, although this is reduced somewhat by hoop stress of the gasket lip. While the two piston rings are designed and adjusted to mechanically perform as similarly as possible, one of them is vented through, so that the seal only occurs on one side and there is no trapped volume between the two rings. As shown in Figure~\ref{fig:flow}, prototypes of the pump relying only on the elastic properties of the plastics to provide radial sealing provided inferior performance, as expected.

To minimize outgassing from the UHMWPE gaskets, they were baked in a N$_2$-purged oven for 200~hours at 93$^{\circ}$C before installation. A 0.003~$\mu$m particulate filter\cite{mott} is installed at the output of the pump to capture any particulate that may be generated. Gas from the pump also passes through the hot zirconium purifiers before returning to the xenon vessel. Purity is measured with a gas purity monitor \cite{gpm} at the output of the purifiers and it is found to be sufficient for the operation of a liquid xenon detector with a $\simeq$20~cm drift distance. It is anticipated that the piston rings will wear over time and may need replacing after several months, although no appreciable wear has been seen after 1800 hours of operation.

\section{Performance characterization}

Two largely identical magnetically coupled piston pumps were built as part of the xenon recirculation system in the EXO-200 experiment. The initial prototype (Pump 1), with 2.5~cm long external and internal magnets and no gasket tensioning mechanism, has been operated with two different sets of PFA piston rings (designated A and B, and differentiated primarily by a slight difference in diameter in an attempt to optimize the sealing simply using the elasticity of the plastics). The production device (Pump 2), designed for higher differential pressure, is made with 5.1~cm long magnets and uses sprung UHMWPE gaskets. Also in this case two sets of gaskets with slightly different diameters (designated C and D) were tried. The pump is typically operated at 14 strokes/min, corresponding to a linear velocity of 8~cm/s in the middle of the stroke.

\begin{figure}
\includegraphics[width=80mm]{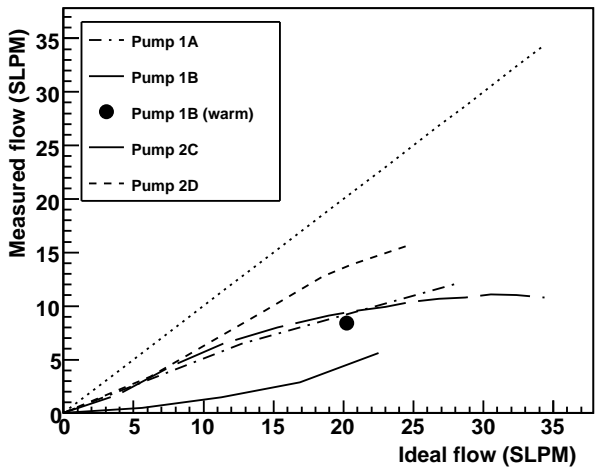}
\caption{The measured xenon flow rate through the two pumps and four total sets of gaskets at inlet pressures of approximately 760~torr. The ideal flow is calculated from the pump speed and the dotted line represents 100\% efficiency. Pump 1B shows a dramatic improvement in efficiency after some running time (about two hours), so an additional point is plotted after allowing the gaskets to warm up. This warmup effect is much less significant for Pump 1A, and negligible for Pump 2. The maximum flow rate in Pump 1 is limited by the coupling force between the magnets, while for Pump 2 magnet decoupling occurs at much higher differential pressure and higher flow rates are possible.}
\label{fig:flow}
\end{figure}

The efficiency of the pumps is investigated by comparing the measured mass flow rate to the nominal one calculated as the volumetric displacement of the piston, as shown in Figure~\ref{fig:flow}. Both imperfect radial seals and back flow though the reed valves degrade the efficiency. The data in Figure~\ref{fig:flow} was taken with inlet pressure near 760~torr, but we have observed increasing efficiency as the inlet pressure is increased. We have also consistently observed higher efficiency when pumping room air in bench tests of the pumps when not installed in EXO-200---an effect that is not understood. 

The performance of Pump 1 with un-sprung PFA gaskets critically depends on the exact tolerance of the gaskets and, in fact, the efficiency of the looser gaskets of 1B substantially improves after a warm-up period (and, presumably, expansion). This is illustrated by the point in Figure~\ref{fig:flow} for Pump 1, resulting from gasket warm-up in the first few hours of operation. Pumping efficiency for Pump 1 also increases with increases in room temperature for the same reason. Tighter gaskets (1A) have better initial performance but they wear out in a few days of continuous pumping (not shown in Figure~\ref{fig:flow}). In the case 1B the efficiency decreased by 80\% after 1500 hours of operation from gasket wear. The sprung gaskets made of UHMWPE provide consistent performance at all times and are durable (no loss of performance has been observed in the first 1800 hours of operation).

For Pump 1 the maximum pumping rate is limited ultimately by the differential pressure at which the piston will decouple from the external magnet. The differential pressure at decoupling was measured to be at least 400 (600) torr for case 1A (1B). The difference between the two gaskets is attributed to additional force due to friction from the tighter fitting A gaskets.

The decoupling regime has not been studied extensively for Pump 2. For the C gaskets, low pumping efficiency at high piston speeds limits the maximum pumping rate. For the D gaskets we have successfully operated with differential pressures up to 750~torr, although this is rarely required in EXO-200 operations. This is sufficient to recirculate xenon at more than 16~SLPM through the EXO-200 recirculation loop with typical inlet pressures of 700--1000~torr.

\section{Conclusion}

We have designed and built an ultra-clean magnetically coupled pump for xenon gas recirculation in EXO-200. The pump, constructed from stainless steel ConFlat\textsuperscript\textregistered\ components with no dynamic seals to the outside, provides a structurally robust alternative to bellows or diaphragm designs. Installed as part of the EXO-200 xenon recirculation system, the pump has demonstrated average gas flow rates up to 16~SLPM with an inlet gas pressure of 700--1000~torr and a differential pressure across the pump up to 750~torr. Piston gasket wear, resulting in reduced performance over time, is still being investigated, but we have operated one set of gaskets for over 1800~hours with no loss of performance, allowing us to operate the experiment as designed with sufficient purity.

% If in two-column mode, this environment will change to single-column format so that long equations can be displayed. 
%\begin{widetext}
%$$\mbox{put long equation here}$$
%\end{widetext}

% Figures should be put into the text as floats. 
% Use the graphics or graphicx packages (distributed with LaTeX2e).
% Put the label that you will use with \ref{} command in the braces of the \label{} command.
%
% \begin{figure}
% \end{figure}

% \begin{table}
% \end{tabular}
% \end{table}

\begin{acknowledgments}
We thank K. Merkle and J. Kirk (Stanford) for the careful machining of parts and Prof. D. DeBra (Stanford) for the suggestion of using UHMWPE gaskets. EXO is supported by DoE and NSF in the United States, NSERC in Canada, FNS in Switzerland and MSE and RFBR (N~10-02-00484) in Russia.
% Put your acknowledgments here.
\end{acknowledgments}


\begin{thebibliography}{18}
\bibitem [{ext()}]{extmag}%
  \BibitemOpen
  \href@noop {} {}\bibinfo {note} {The external magnet is made from a stack of
  four identical magnets (Applied Magnets model NR023-1).}\BibitemShut {Stop}%
\bibitem [{bra()}]{brass}%
  \BibitemOpen
  \href@noop {} {}\bibinfo {note} {A separate test using brass was
  unsuccessful.}\BibitemShut {Stop}%
\end{thebibliography}
\end{document}